\documentclass[conference]{IEEEtran}
\IEEEoverridecommandlockouts
\usepackage[left=1.7cm,right=1.6cm,top=1.9cm,bottom=2.3cm]{geometry}
\usepackage[final]{graphicx}
\usepackage{multicol}
\graphicspath{{Images//}} 

\usepackage{hyperref}
\usepackage{psfrag}
\usepackage{epstopdf}
\usepackage{amsfonts}
\usepackage{mathrsfs}
\usepackage{pifont}
\usepackage{verbatim}
\usepackage{upgreek}
\usepackage{color}
\usepackage[usenames,dvipsnames]{xcolor}
\usepackage{algorithm}
\usepackage{mdwmath}
\usepackage{mdwtab}
\usepackage{amssymb,amsmath}
\usepackage{amsmath}
\usepackage{amsthm}
\usepackage{epstopdf}
\usepackage{cite}
\usepackage{tikz}
\usepackage{scalefnt}
\usepackage{subfigure}
\usepackage{MnSymbol}
\usepackage{setspace}
\usepackage{mathtools}
\usepackage{psfrag}
\usepackage{siunitx}

\usepackage[font=scriptsize]{caption}
\usepackage{algpseudocode}

\UseRawInputEncoding

\algrenewcommand\algorithmicforall{\textbf{foreach}}
\algrenewcommand\algorithmicindent{.8em}

\label{newcommand}

\label{English Chars}

\label{Roman Chars}

\label{Theorems}

\theoremstyle{remark}

\begin{document}
\label{title}
\title{D-Band RIS as a Reflect Array:\\ Characterization and Hardware Impairments Study}

\author{
	\IEEEauthorblockN{Ehsan Tohidi\IEEEauthorrefmark{1}\IEEEauthorrefmark{2}, Robert St\"ocker\IEEEauthorrefmark{3}, Julia-Marie K\"oszegi\IEEEauthorrefmark{3}, and Slawomir Sta\'nczak\IEEEauthorrefmark{1}\IEEEauthorrefmark{2}
		}\\
	\IEEEauthorblockA{\IEEEauthorrefmark{1}%
		Fraunhofer HHI, 
\IEEEauthorrefmark{2} Technische Universit\"at Berlin,
\IEEEauthorrefmark{3} Fraunhofer IZM,\\
Email: \{ehsan.tohidi, slawomir.stanczak\}@hhi.fraunhofer.de,\\ \{robert.stoecker, julia-marie.koeszegi\}@izm.fraunhofer.de}\,\\
\thanks{The authors acknowledge the financial support by the Federal Ministry of Education and Research of Germany in the programme of ``Souver\"an. Digital. Vernetzt.'' Joint project 6G-RIC, project identification number: 16KISK020K and 16KISK030.}
}

\maketitle

\begin{abstract}

 Reflecting intelligent surface (RIS) has emerged as a promising technology for enhancing wireless communication performance and enabling new applications in 6G networks with potentially low energy consumption and hardware complexity thanks to their passive nature. Despite the significant growth of the literature on RIS in recent years, covering various aspects of this technology, challenges and issues regarding the practical implementation of RIS have been less addressed. This issue is even more severe at D-band frequencies due to the inherent challenges. This paper aims to connect the requirements and aspirations from the link-level side to the actually achievable RIS hardware with the focus on the various models for RIS. In order to obtain a realistic hardware scenario while maintaining a manageable parameter set, a static reflect array with similar reflection behavior as the reconfigurable one is employed. The results of the study enable an improved RIS design process due to more realistic models. The special focus of this paper is on the hardware impairments, in particular, $(i)$ the effect of specular reflection, and $(ii)$ the beam squint effect. 
\end{abstract}

\section{Introduction}
The sixth generation (6G) represents a significant leap forward in wireless communications, potentially enabling new applications and services with high quality of service (QoS) requirements and, at the same time, reducing the network energy consumption and hardware costs. Reflecting intelligent surface (RIS) is a promising technology that can help meet the ambitious goals of 6G by enhancing the performance of wireless networks in a variety of ways. By manipulating the reflections of electromagnetic waves, RIS can create additional wireless paths, amplify weak signals, and mitigate interference. Due to the passive nature of RISs, they are considered as a technology with low energy consumption and low hardware cost \cite{wu2019intelligent, pan2021reconfigurable, wu2021intelligent, Tohidi2023icc}.

Coverage extension, sensing, channel matrix rank enhancement, and security are of the known promising RIS use cases. Moreover, RISs can be used as reflect arrays in order to compensate for the limited degrees of freedom (in terms of the number of antennas and beamforming capabilities) at the transmitter. This means that we can move the beamforming function from the radio endpoints of the environment, consequently relaxing the constraints on the transmitter side.

\subsection{Literature Review}
While the literature on RIS has grown significantly in recent years, there are still some shortcomings that need to be addressed. One of the main issues is the rare practical implementation and demonstration of the proposed theoretical models and algorithms. Most of the existing research on RIS has focused on simulation-based studies and theoretical analysis, which may not fully capture the practical challenges and limitations of real-world deployment.

Authors in \cite{arun2020rfocus}, use passive antennas as a reflect array in order to relax the size constraint of the radio transmitter. Moreover, in \cite{pei2021ris},
a RIS prototype consisting of 1100 controllable elements working at 5.8 GHz band is presented which enables a link with a high data rate between the transmitter and receiver, while the direct link is blocked. The path loss of the RIS-assisted wideband channel based on the field channel measurements at 26 GHz in an indoor corridor scenario is analyzed in \cite{10008687}. The proposed metasurface design in \cite{zhu2014dynamic} enables dynamic wave control by  producing arbitrary reflection magnitude and phase. Also, authors in \cite{tang2020wireless} have developed free-space path loss models for RIS-assisted wireless communications by studying the physics and electromagnetic nature of RISs. The proposed models in \cite{tang2020wireless} are first validated through simulation results and then, three fabricated RISs (metasurfaces) are utilized to further corroborate the theoretical findings through experimental measurements conducted in a microwave anechoic chamber. However, the study remains limited to 4.25 GHz and 10.5 GHz in the three manufactured RISs.

To the authors' knowledge, there is no RIS for D-Band available commercially or investigated with hardware in the literature. However, a good starting point for the investigation of RIS is reflect array (RA) antennas which do exist and are studied in the D-band as shown by the following references. In \cite{Zhao2018}, an RA is proposed at $\SI{140}{\giga\hertz}$ in PCB technology. The size of the RA is $\SI{100}{\milli\meter}^2$ with a half-wavelength lattice. An RA antenna at $\SI{142}{\giga\hertz}$ with a size of $\SI{4.3}{\milli\meter}^2$ is investigated in \cite{Yang2022}. A gain of $\SI{23.75}{\decibel}i$ is achieved. In \cite{Jiajia2021RA}, an RA at $\SI{115}{\giga\hertz}$ is presented. This RA is realized in PCB technology and achieves a gain of $\SI{18}{\decibel}i$. An RA at $\SI{120}{\giga\hertz}$ with a diameter of $\SI{138}{\milli\meter}$ realized with lithography technology is presented in \cite{Tamminen2013RA} while authors in \cite{Peng2022RA}, propose an RA at $\SI{110}{\giga\hertz}$ with a gain of $\SI{22.3}{\decibel}i$.


\subsection{Contributions}
This work is considered a proof of concept to close the gap between theory on the link and network side and practice on the hardware side. The goals of this paper are two-fold: $(i)$ characterize an RA as a model for an individual RIS state and identify required adjustments to have more realistic models for the achievable gain and radiation properties and $(ii)$ identify hardware impairments, in particular, specular reflection and beam squint effect, which majorly contribute to a performance loss of RISs in real scenarios.
We use a fabricated RA without the extra features of a RIS, e.g., the control system, which are irrelevant to this study. We only focus on the reflection properties of the designed surface. The details of the RA design, simulation and experimental characterization are provided in \cite{Stoecker2023}. Ref. \cite{Stoecker2023} has a strong focus on the hardware implementation of the RA at the carrier frequency $f_0\,=\,\SI{150}{\giga\hertz}$. The present study builds on the reference and provides the answer to the question of how the described radiation pattern will influence signal propagation once the RA is integrated into a network. Therefore, the present study will act as an enabler for later RIS implementations.

\section{System Model}

\begin{figure}
    \centering
    \includegraphics[scale=0.5]{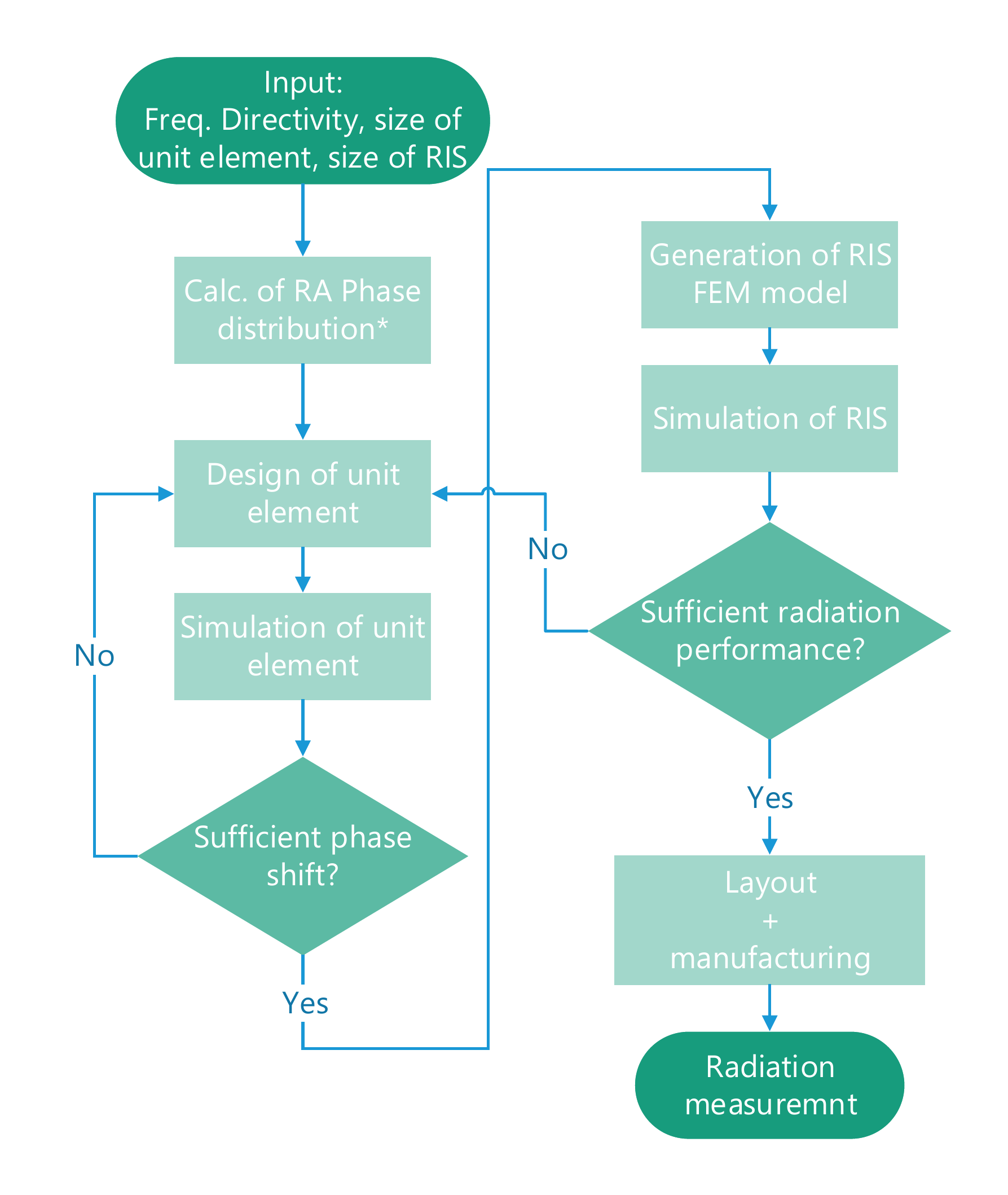}
    \caption{Flow chart for RIS hardware synthesis. FEM stands for Finite Element Method. * Marks the box which is specific for RA antenna design. In the RIS workflow the distribution is not fixed but controllable.}
    \label{fig_design_FlowChart}
\end{figure}

In the following section, we present the considered RIS structure and explain the proposed design procedure. The design flow for a RIS is outlined in Fig. \ref{fig_design_FlowChart} and can be divided into two steps: The first step is the design of the unit element which provides the local phase shift for the reflected wave. In the second step, the complete RIS is synthesized and simulated. In a RIS, the unit elements will all have similar shapes and the local phase shift is (electronically) adjusted. In contrast, the unit elements of RA (as in Ref. \cite{Stoecker2023}) vary in their shape to realize the required phase shift. In the following sections, the RIS design steps are described in further detail.


\subsection{RIS Structure}
We consider a large square static surface designed at the carrier frequency $f_0\,=\,\SI{150}{\giga\hertz}$ to act as an RA, placed in the $xy$-plane. The RIS is composed of unit elements with $\lambda_0/4$ spacing, both vertically and horizontally, with $\lambda_0$ being the wavelength at the carrier frequency. The incident angle was chosen to be at $\varphi_0\,=\,\SI{0}{\degree}$ and $\vartheta_0\,=\,\SI{60}{\degree}$. The reflecting angle was chosen at $\SI{0}{\degree}$ in both the $\varphi$-plane and the $\vartheta$-plane (broadside). The RA is designed to receive from a $\SI{19.3}{\milli\meter}$ distance (within near-field distance). The large receiving angle was chosen to generate realistic data for the RIS use case.

\subsection{Design of the Unit Element}



The unit element defines the local phase shift at each lattice position of the RIS. The design of unit elements is the base of the RIS design. For a static RA antenna, the local phase shift can be obtained by changing the geometry of each unit element like variable rotation, variable size, or variable delay lines. The various approaches are described in \cite{Nayeri2018book}. In all cases, the hardware is designed for fixed angles of incidence and reflection. For a RIS, the adjustable phase shift of each element requires more advanced hardware approaches including changing material properties or changing capacitance. However, the target operation within the D-Band significantly limits the options. In the present study, an RA antenna based on PCB technology is chosen to represent one state of a RIS because its implementation is closest to possible RIS hardware. The structure of the optimized unit elements, the realized phase shift, and the resulting reflection coefficient are in detail described and are displayed in figures 1-3 in \cite{Stoecker2023}.

\subsection{Design of RIS} \label{sec:DesignRIS}
Based on the simulation of unit elements, the RIS is synthesized with a Python script as indicated in the right column of the design flow chart in Fig. \ref{fig_design_FlowChart}. The phase distribution can be calculated according to \eqref{eqn_phase_dristibution} \cite{Nayeri2018book}:
\begin{align}\label{eqn_phase_dristibution}
    \varphi_{RA} = k_0 \left( x_i \sin \vartheta_0 \cos \varphi_0 + y_i \sin \vartheta_0 \sin \varphi_0 \right)
\end{align}
with $\varphi_{RA}$ being the required phase shift at each element of position $x_i$ and $y_i$, and $k_0$ is the wavenumber at the carrier frequency.

\begin{figure}
    \centering
    \includegraphics[width=0.49\textwidth]{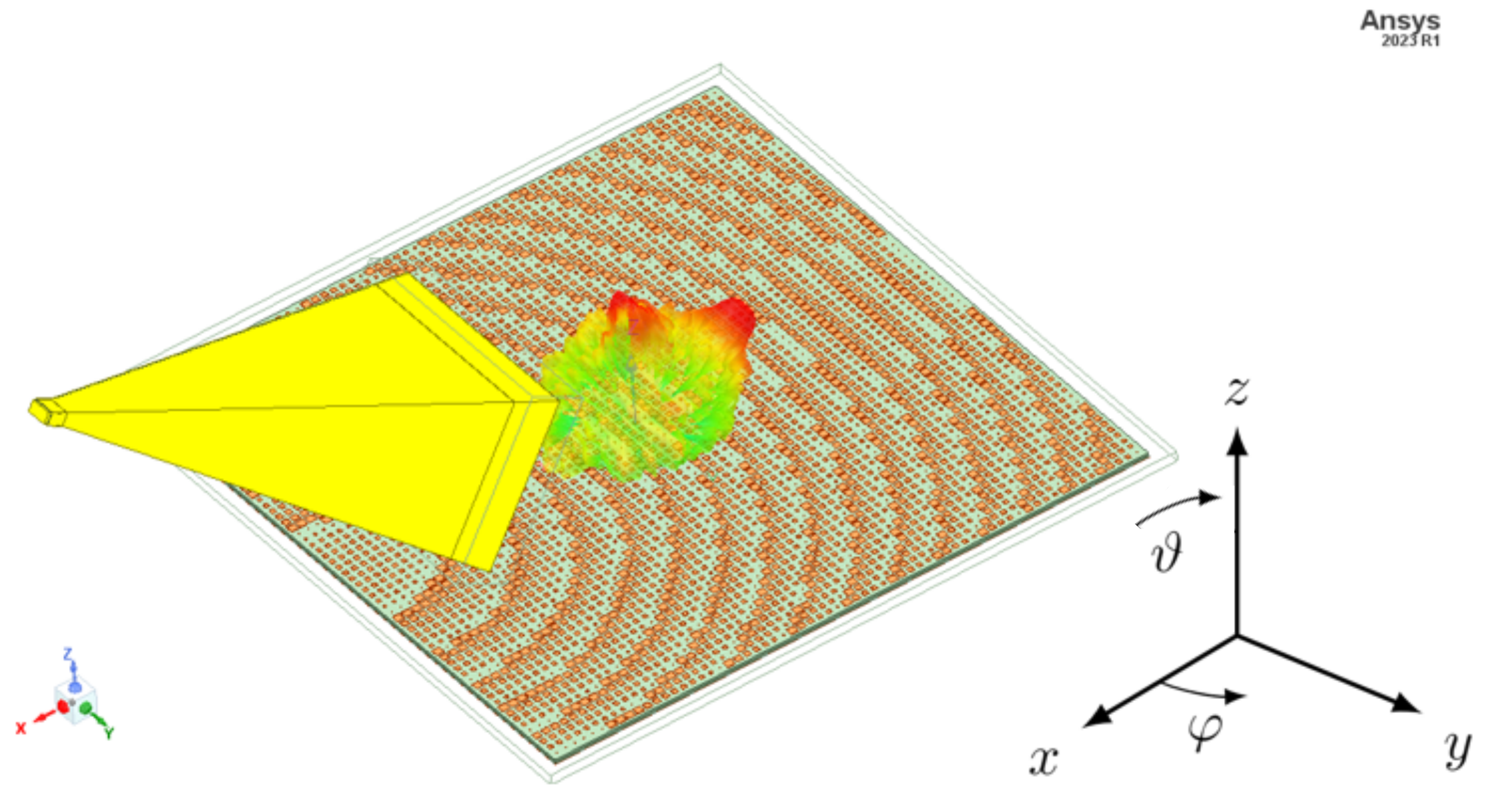}
    \caption{The simulation setup including the RIS, the feeding horn antenna, and the radiation pattern \cite{Stoecker2023}.}
    \label{fig_Model_and_pattern}
\end{figure}

Fig. \ref{fig_Model_and_pattern} illustrates the HFSS model of the static RIS in the $xy$-plane with the feed horn antenna and a radiation pattern. The feed horn antenna points from a distance of $\SI{19.3}{\milli\meter}$ with an angle of arrival of $\vartheta_0 = \SI{-60}{\degree}$ and $\varphi_0 = \SI{0}{\degree}$. The horn antenna is a standard gain horn antenna with a gain of $\SI{25}{\decibel}i$. Two RIS sizes are considered in this paper: a larger surface of $\SI{36}{\milli\meter} \times \SI{36}{\milli\meter}$, and a smaller one of $\SI{18}{\milli\meter} \times \SI{18}{\milli\meter}$. The comparison of different sizes is a helpful tool to analyze the hardware impairments and their scaling.


\section{RIS Radiation Pattern and Analysis of Impairments}
In the interest of space limit and due to the main function of a RIS as an RA in this paper, we present simulations regarding reflection radiation pattern, RIS gain, and the two major hardware impairments namely specular reflection and beam squint effects which were confirmed by experiments \cite{Stoecker2023}. 

\subsection{Simulation Results}
Fig. \ref{fig_Directivity_RA_150G_lambda_4tel_stacked} displays the radiation pattern for the larger RIS of size $\SI{36}{\milli\meter} \times \SI{36}{\milli\meter}$. Fig.
\ref{fig_gain_RA_150G_lambda_4tel_stacked_half} plots the radiation pattern for the smaller RIS of size $\SI{18}{\milli\meter} \times \SI{18}{\milli\meter}$. 

To investigate the effect of frequency deviation from the nominal frequency, Figs. \ref{fig:phi_beam_squint} and \ref{fig:theta_beam_squint} depict the radiation pattern at the three frequencies $\SI{140}{\giga\hertz}$, $\SI{150}{\giga\hertz}$, and $\SI{160}{\giga\hertz}$ for the larger surface of size $\SI{36}{\milli\meter}\times\SI{36}{\milli\meter}$ in $\varphi$- and $\vartheta$-planes.

\subsection{Analysis: Model Verification and Hardware Impairments}
Up to this point, the RIS was discussed purely from the hardware design point of view with the aim to realize the maximum directivity at a required frequency with the available technology (material properties, geometrical parameters). In other words, maximize the RIS performance within the defined parameter space. However, the RIS needs to be integrated into a network and the hardware models need to be accompanied by proper models for signal propagation. Therefore we will shift the point of view in the following to the required description in a link analysis context.

\subsubsection{Reflection Radiation Pattern}
As depicted in Figs. \ref{fig_Directivity_RA_150G_lambda_4tel_stacked} and \ref{fig_gain_RA_150G_lambda_4tel_stacked_half} the main lobes occur at $\SI{0}{\degree}$ at the nominal frequency of $\SI{150}{\giga\hertz}$, consistent with the designed model for both simulated sizes of RIS. The beamwidth for the larger RIS is $\SI{12}{\degree}$ and $\SI{6}{\degree}$ in $\varphi$- and $\vartheta$-planes, respectively. For the smaller RIS, a larger beamwidth in the main lobe can be observed with $\SI{14}{\degree}$ in both the $\varphi$- and $\vartheta$-planes.

\subsubsection{Gain}
\begin{figure}
    \centering   \includegraphics{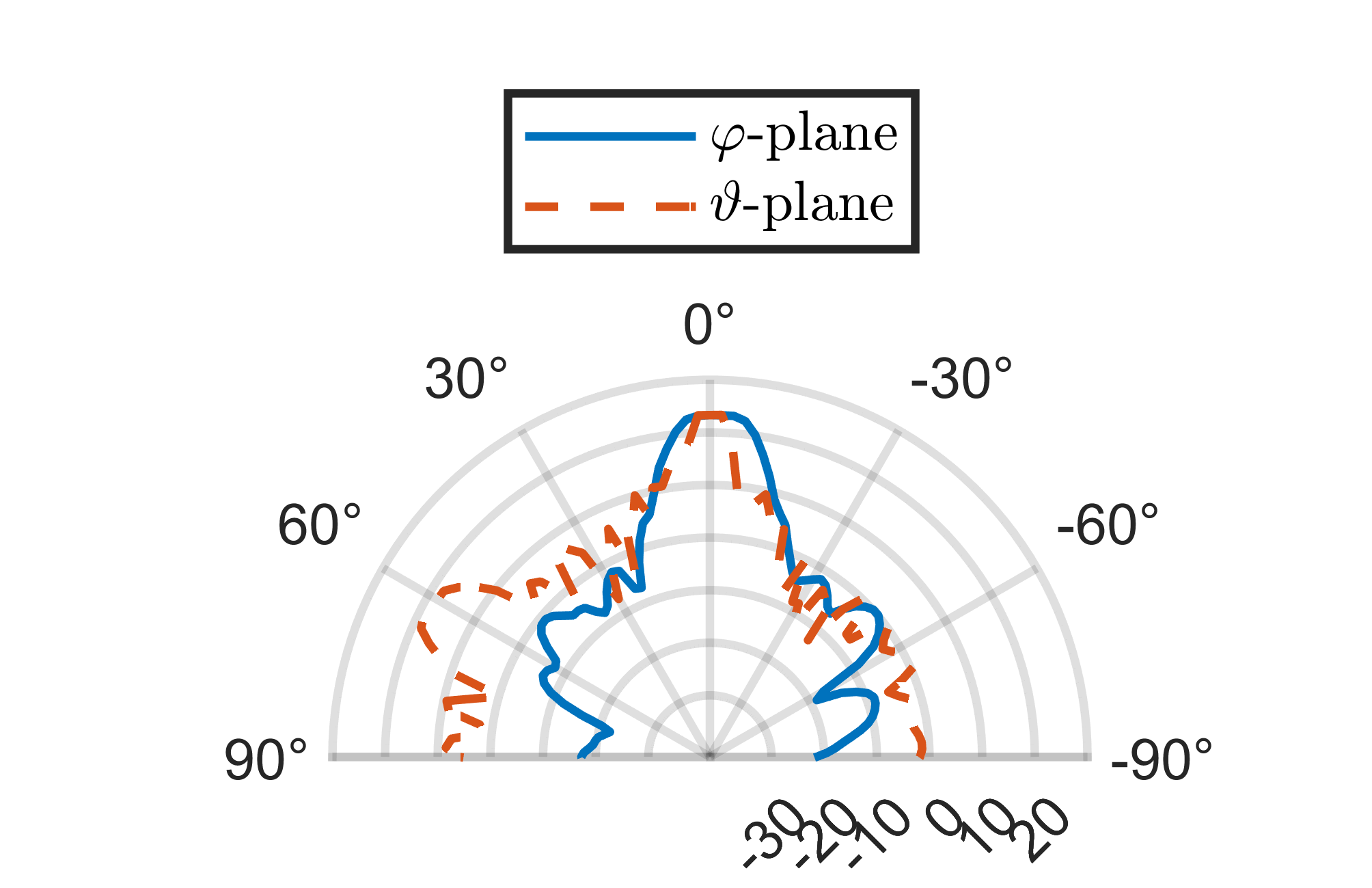}
    \caption{Radiation pattern of the $\SI{36}{\milli\meter}\times\SI{36}{\milli\meter}$ RIS at 150 GHz.}
    \label{fig_Directivity_RA_150G_lambda_4tel_stacked}
\end{figure}

\begin{figure}
    \centering    \includegraphics{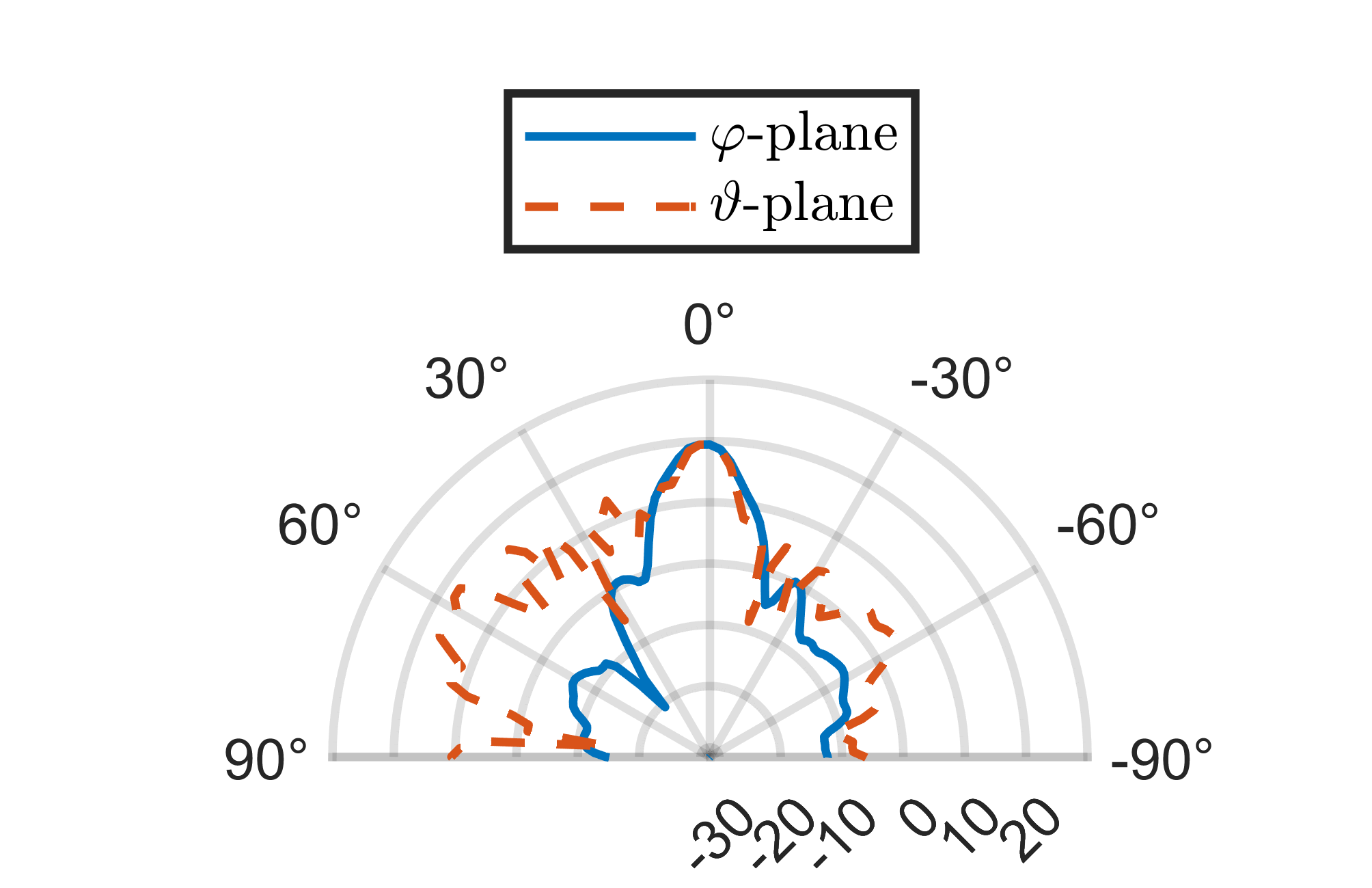}
    \caption{Radiation pattern of the $\SI{18}{\milli\meter}\times\SI{18}{\milli\meter}$ RIS at 150 GHz.}
    \label{fig_gain_RA_150G_lambda_4tel_stacked_half}
\end{figure}

The gain of a RIS can be approximated by \cite{Nayeri2018book}:
\begin{align}\label{eqn_gain}
    G = \eta D = \eta \frac{4\pi}{\lambda^2}A_{aperture}
\end{align}
with $A_{aperture}$ as effective aperture of the RIS and $\eta$ is the RIS efficiency defined in \cite{Nayeri2018book}.
Ideally, the effective aperture in \eqref{eqn_gain} is equal to the physical aperture of the RIS. However, it is determined by its behavior when it interacts with electromagnetic waves. Given the physical aperture as the input to \eqref{eqn_gain} and neglecting the efficiency parameter (i.e., $\eta=1$), we expect gains of $\SI{36}{\decibel}$ and $\SI{30}{\decibel}$ for the large and small RIS, respectively. However, the simulations result in gain values of $\SI{23.3}{\decibel}$ and $\SI{18.6}{\decibel}$, respectively (Fig. \ref{fig_Directivity_RA_150G_lambda_4tel_stacked} and \ref{fig_gain_RA_150G_lambda_4tel_stacked_half}). For the difference, we need to investigate the model impairments some included in the following.


\subsubsection{Specular Reflection}
Due to the ground plane of the RIS with a geometric size considerably larger than the wavelength, in reality, besides the designed anomalous reflection, we experience specular reflection which causes both power loss and unintentional interference. As depicted in Figs. \ref{fig_Directivity_RA_150G_lambda_4tel_stacked} and \ref{fig_gain_RA_150G_lambda_4tel_stacked_half}, the specular reflection gains are $\SI{18}{\decibel}$ and $\SI{17.8}{\decibel}$ for the larger and smaller RISs, respectively. Considering the gains of each surface at the intended direction (i.e., $\vartheta=0$), specular reflection contributes almost $\SI{1.1}{\decibel}$/$\SI{3}{\decibel}$ for the larger /smaller surfaces. On the other hand, this can lead to an average signal-to-interference ratio (SIR) of $\SI{5.3}{\decibel}$ and $\SI{0.2}{\decibel}$ for the larger and smaller surfaces, respectively. Unless we find a use case that utilizes the specular reflection, and such a use case is not clear at the time being, this impairment can potentially cause a large degradation in the system performance when RIS is integrated into the wireless network. In other words, besides taking the loss into account in the calculations of the link power budget, the resulting interference needs to be considered. Furthermore, future implementations of RIS need to prioritize the suppression of the specular reflection.

\begin{figure}
    \centering
    \includegraphics{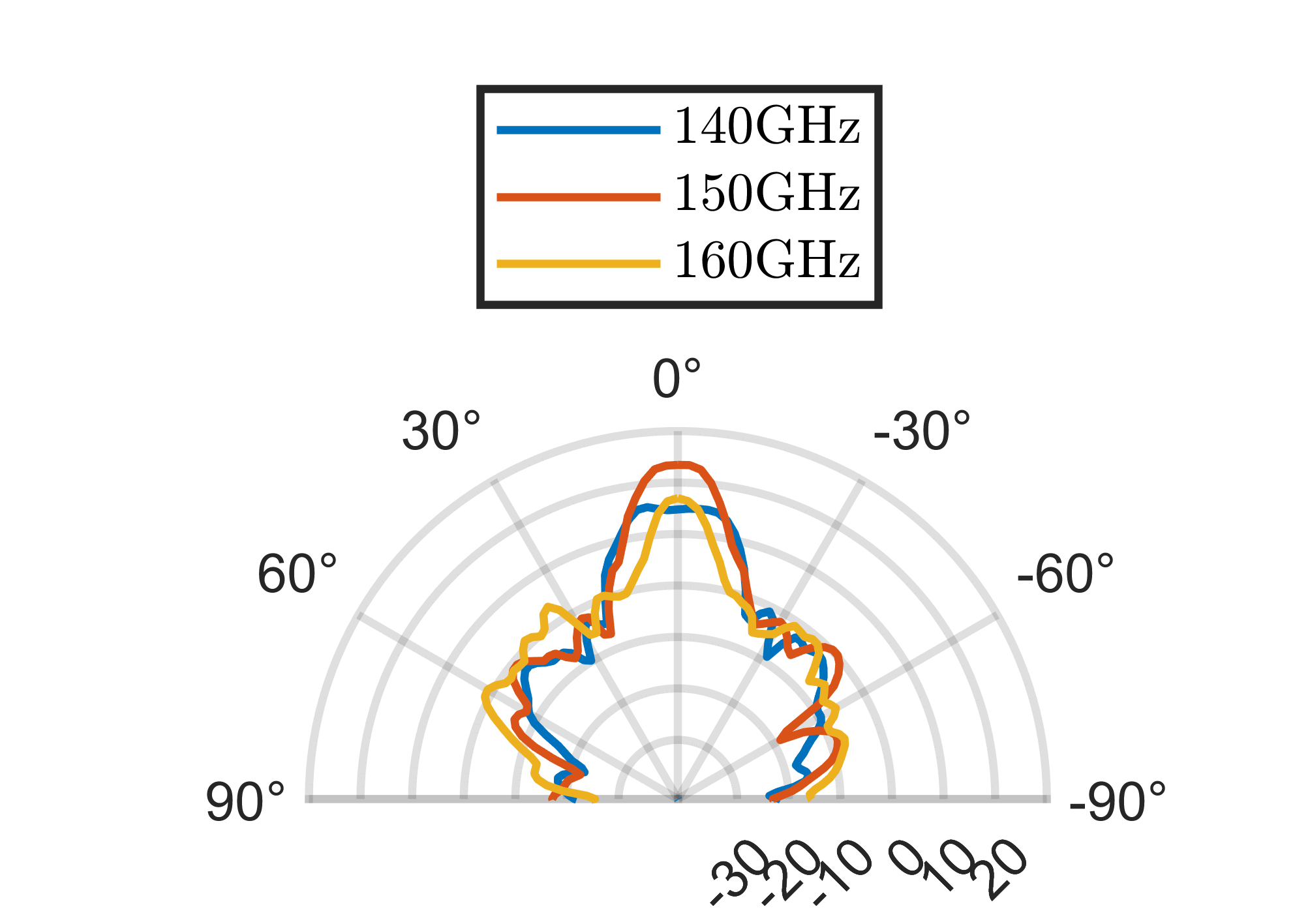}
    \caption{Beam-squint in the $\varphi$-plane ($\SI{36}{\milli\meter}\times\SI{36}{\milli\meter}$ RIS).}
    \label{fig:phi_beam_squint}
\end{figure}

\begin{figure}
    \centering
    \includegraphics{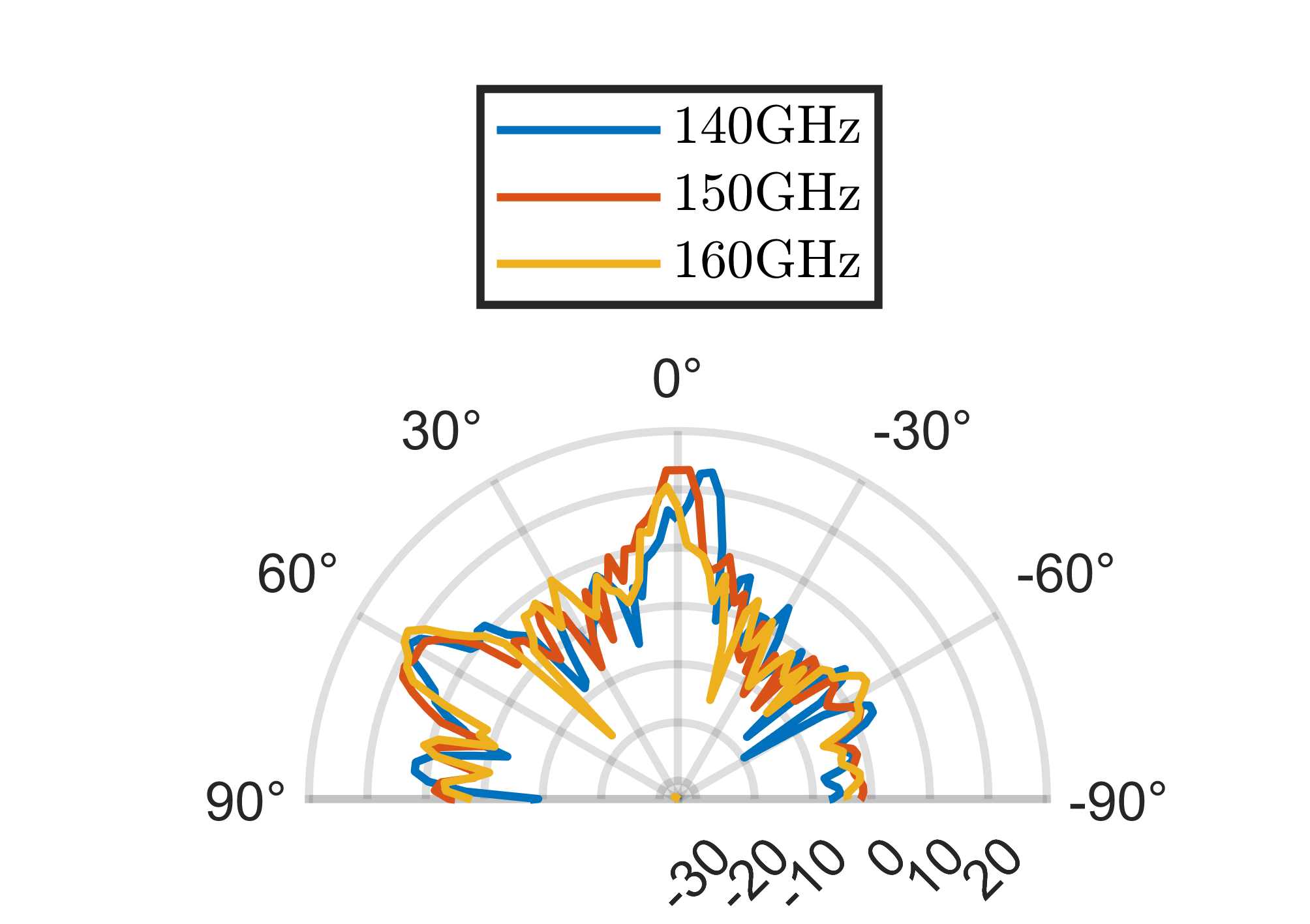}
    \caption{Beam-squint in the $\vartheta$-plane ($\SI{36}{\milli\meter}\times\SI{36}{\milli\meter}$ RIS).}
    \label{fig:theta_beam_squint}
\end{figure}

\subsubsection{Beam Squint Effect}
The unit elements of the RA are designed for a specific frequency, similar to analog beamforming where the sets of phase shifter values are designed at a specific frequency (i.e., the carrier frequency). However, the RIS is subjected to all frequencies within the transmission bandwidth. While phase shifters are a good approximation to ideal time shifters for narrowband transmission, the approximation becomes less accurate for wideband transmission when the angle of arrival or departure is not broadside, as the required phase shifts become frequency-dependent. As a result, beams for frequencies other than the carrier frequency experience beam squint, where their direction varies with frequency. The performance for wideband transmission is especially important in the analysis of D-band hardware since the high bandwidth is one major motivation for D-band applications. 

As shown previously, \eqref{eqn_phase_dristibution} calculates the phase shifts for the considered incident angle (i.e., $\vartheta = \SI{-60}{\degree}$ and $\varphi = \SI{0}{\degree}$) and reflection angle (i.e., $\vartheta = \SI{0}{\degree}$ and $\varphi = \SI{0}{\degree}$) at the nominal frequency (i.e., $f_0 = \SI{150}{\giga\hertz}$ with the corresponding wavelength $\lambda_0$). To see the beam deviation at another frequency $f_1$ with the corresponding wavelength $\lambda_1$, a straightforward calculation leads to
\begin{equation}
\begin{aligned}
\vartheta_1 = \arcsin\left(\frac{\lambda_1}{\lambda_0}\sin \vartheta_0\right)
\end{aligned}
\end{equation}
with $\vartheta_1$ being the new shift between incident and reflection angles in the $\vartheta$-plane. We have conducted simulations as depicted in Figs. \ref{fig:phi_beam_squint} and \ref{fig:theta_beam_squint} for two additional frequencies ($\SI{140}{\giga\hertz}$ and $\SI{160}{\giga\hertz}$). As mentioned earlier, there is no beam squint effect in the $\varphi$-plane since the shift in $\varphi$-plane is designed to be $0$. The absence of the effect is verified by the simulation results in Fig. \ref{fig:phi_beam_squint} and by measurements \cite{Stoecker2023}. In the $\vartheta$-plane, the frequencies of $\SI{140}{\giga\hertz}$ and $\SI{160}{\giga\hertz}$ result in new shifts between incident and reflection angle equal to $\SI{-68.1}{\degree}$ and $\SI{-54.3}{\degree}$, respectively. These shifts in $\vartheta$-plane result in peaks at $\SI{-8.1}{\degree}$ and $\SI{5.7}{\degree}$ in the radiation pattern for the fixed incident angle at $\vartheta = \SI{-60}{\degree}$. The plots with the simulation results in Fig. \ref{fig:theta_beam_squint} indicate values of $\SI{-6}{\degree}$ and $\SI{2}{\degree}$, respectively. Although the theoretical expressions are able to predict the beam deviations up to good precision, still, the hardware impairments and unknown parameters cause slight differences between the calculations and the measurements.

\subsubsection{Further Impairments}
Although we have investigated two major hardware impairments, there are further known sources of impairments of noticeable effect in the near-field region, that require further studies:
\begin{itemize}
    \item The \textit{spillover efficiency} measures the amount of radiation from the feed antenna that is reflected by the surface. Due to the finite size of the RIS, some of the radiation from the feed antenna will not be received by the RIS and thus, not be reflected.
    \item The \textit{point source assumption} of feed antenna: In designing the phase shifts according to \eqref{eqn_gain}, the implicit assumption is the feed antenna being a point source. Clearly, this is not the case in practice which leads to a slight mismatch of the phase shifts.
    \item \textit{Amplitude tapering}: This phenomenon is addressing the different power levels received at different unit elements of the RIS which can be potentially noticeable in the near-field region. Amplitude tapering leads to a smaller effective aperture typically due to two reasons in this context: First, the difference in distances between the feed antenna and the RIS with subsequently different path losses, and second, the RIS being illuminated from different parts of the radiation pattern of the feed antenna. The combination of the two effects contributes to the amplitude tapering which is dependent on the size of the RIS. The larger the surface, the higher the amplitude tapering, which leads to higher losses. In other words, one cannot arbitrarily increase the size of the RIS, assuming the same growth in the achieved gain.
\end{itemize}
There is still a need for further studies in order to incorporate the losses corresponding to the impairments into the link budget models and network calculations. For any next step, the result of the present study will guide the design process, e.g. by integrating the SIR by specular reflection and the maximum permitted beam squint into the input parameters for the design and optimization process (Fig. \ref{fig_design_FlowChart}).

\section{Outlook and Conclusions}
The paper aims to foster the next steps in RIS development. 
We use a lower-cost RA as a one-state model for a RIS. We attempt to gain insight into a realistic RIS hardware performance and identify some of the major hardware impairments that impact the performance when integrated into a wireless communication network. 
Full-wave simulation results verify the designed model, particularly in terms of the designed angle of incidence, reflection, and near-field distance. The specular reflection is identified as one important contributor to power loss and can in addition cause unintentional interference. The second investigated impairment is the beam squint. Its effectivity is a function of field of view, relative bandwidth, and the size of the surface, which are potentially quite large values in this context and makes the topic more crucial for RIS modeling. These two impairments are of utmost importance with regards to network integration from two aspects: $(i)$ the decrease in link budget and consequently reduced communication range and $(ii)$ interfering with other users in the network. Furthermore, both are present in near as well as far-field distances to the transmitter.

Following up on the successful simulation results and findings in the present study, prototyping a RIS with an improved design is planned. Furthermore, the utilized RA will be integrated into field measurements with a communication setup to further investigate RIS performance in a real environment at D-band frequencies. Moreover, additional use cases including sensing will be considered for future application since the better angular resolution of the RIS due to the larger aperture compared to the feed antenna provides a better sensing resolution.

%


\ifCLASSOPTIONcaptionsoff
\newpage
\fi
\bibliographystyle{ieeetr}
\bibliography{ref}

\end{document}